\documentclass[12pt]{iopart}
\usepackage[utf8]{inputenc}
\usepackage[T2A]{fontenc}

\usepackage{graphicx}
\begin{document}

\title{Effect of Stone-Wales defects on the thermal conductivity of graphene}

\author{S.E. Krasavin and V.A. Osipov}

\ead{krasavin@theor.jinr.ru and osipov@theor.jinr.ru}
\address{Bogoliubov Laboratory of Theoretical Physics, Joint
Institute for Nuclear Research, 141980 Dubna, Moscow region,
Russia}
\vspace{10pt}
\begin{indented}
\item[]June 2015
\end{indented}

\begin{abstract}
The problem of phonon scattering by strain fields
caused by Stone-Wales (SW) defects in graphene is studied in the framework of
the deformation potential approach. An explicit form of the phonon mean
free path due to phonon-SW scattering is obtained within the Born approximation.
The mean free path demonstrates a specific $q$-dependence varying as $q^{-3}$ at low wavevectors and taking a constant value at large $q$. The thermal conductivity of graphene nanoribbons (GNRs) is calculated with the three-phonon umklapp,
SW and rough edge scatterings taken into account. A pronounced decrease of the thermal conductivity due to SW defects is found at low temperatures whereas at room temperatures and above the phonon-phonon umklapp scattering becomes dominant. A comparison with the case of vacancy defects shows that they play more important role in the reduction of the thermal conductivity in GNRs over a wide temperature range.

\end{abstract}



\maketitle

Structural defects largely affect the thermal properties of graphene and graphene nanoribbons (GNRs), thus affecting the possible thermoelectric applications of graphene-based nanodevices. The most important structural defects experimentally observed in graphene are single and double
vacancies, Stone-Wales (SW) defects, grain boundaries, and
reconstructed defect structures. In this letter, we focus our consideration on the SW defects whose role in the heat transport in graphene is not yet well understood. Studies based on molecular dynamics (MD) demonstrated that SW defects can significantly affect the thermal conductivity of graphene and GNRs within a wide temperature range~\cite{Haskins,Hao,Ng,Yeo}. In particular, it has been found that the presence of SW defects can decrease the thermal conductivity in the temperature range 100-600 K by more than 80 $\%$ as defect densities are increased to 10 $\%$ coverage~\cite{Yeo}. The sharpest decrease was found at low density of SW defects while at high densities less variations in thermal conductivity across a wide range of temperatures were observed~\cite{Ng,Yeo}.

It is interesting that other point-like atomistic modifications show a similar reduction in thermal conductivity of graphene and GNRs (see, e.g.,~\cite{Haskins,pop}). For example, it was found that an increasing number of vacancies led to a decrease of almost 50 $\%$ in the thermal conductivity of GNRs at room temperature~\cite{Hu}. Unlike vacancies, SW defects are topological in their nature and one could expect that this difference will result in some specific features of phonon scattering. It is rather difficult to illuminate these peculiarities in studies based on molecular dynamics because they are limited in the range of sizes that could be examined. The more suitable way is to use the phonon Boltzmann transport equation in the relaxation time approximation. This approach was effective in the description of the thermal conductivity of graphene and GNRs with edge roughness, three-phonon and isotope scattering taken into account~\cite{nika,aks}. In order to estimate the contribution from the SW defects one has to determine the phonon mean free path due to phonon-SW scattering.

Our goal is to calculate the SW-induced phonon mean free path which
would allow us to take into account consistently all main scatterers. In particular, this gives a possibility to compare the role of SW and vacancy scattering.
We consider the canonical SW defects consisting of two pairs of five and seven-membered rings
forming a rhomboid structure (see Fig. 1).
\begin{figure} [t]
\begin{center}
\includegraphics [width=6 cm]{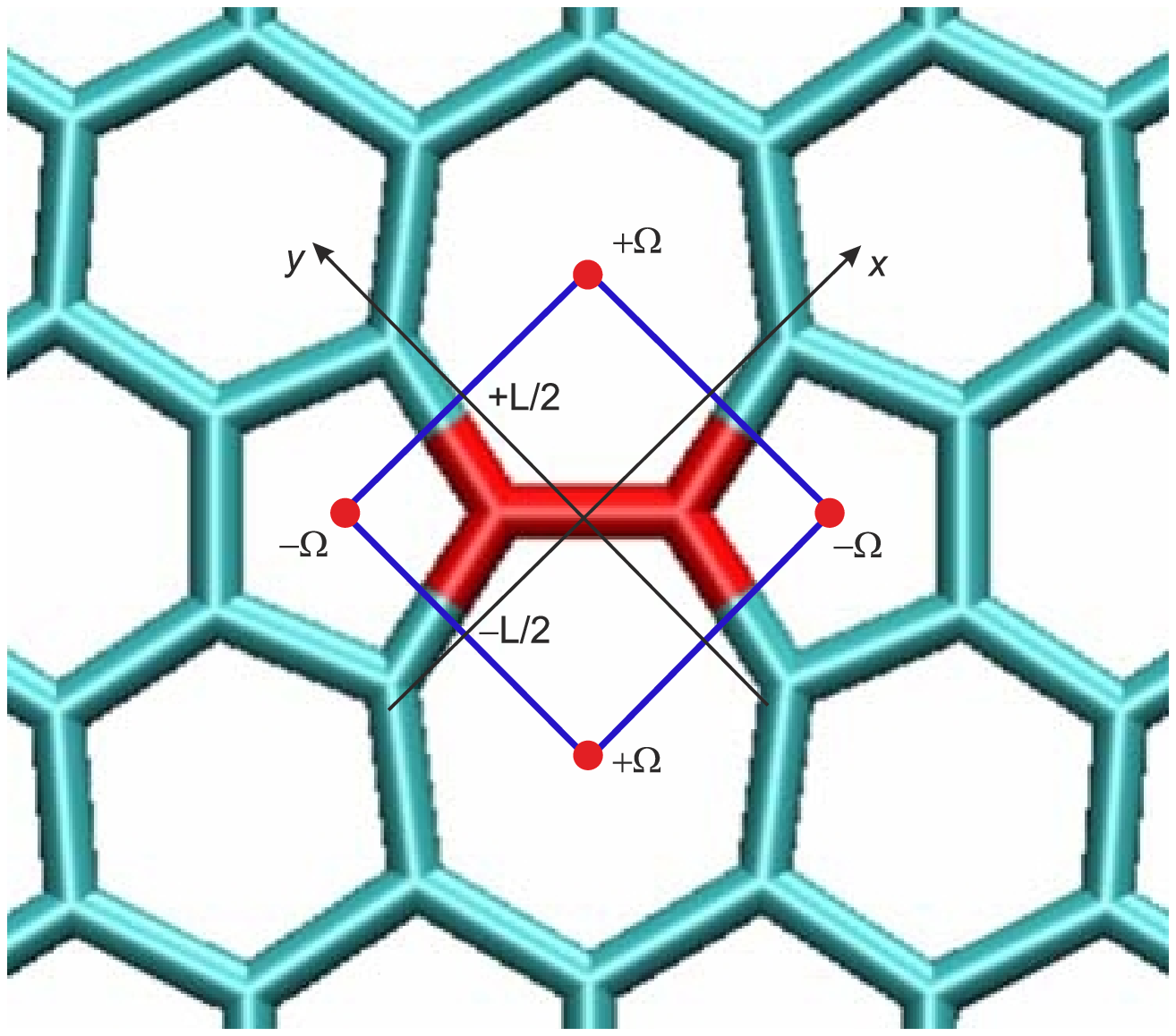} \hspace{2cm}
\includegraphics [width=6 cm]{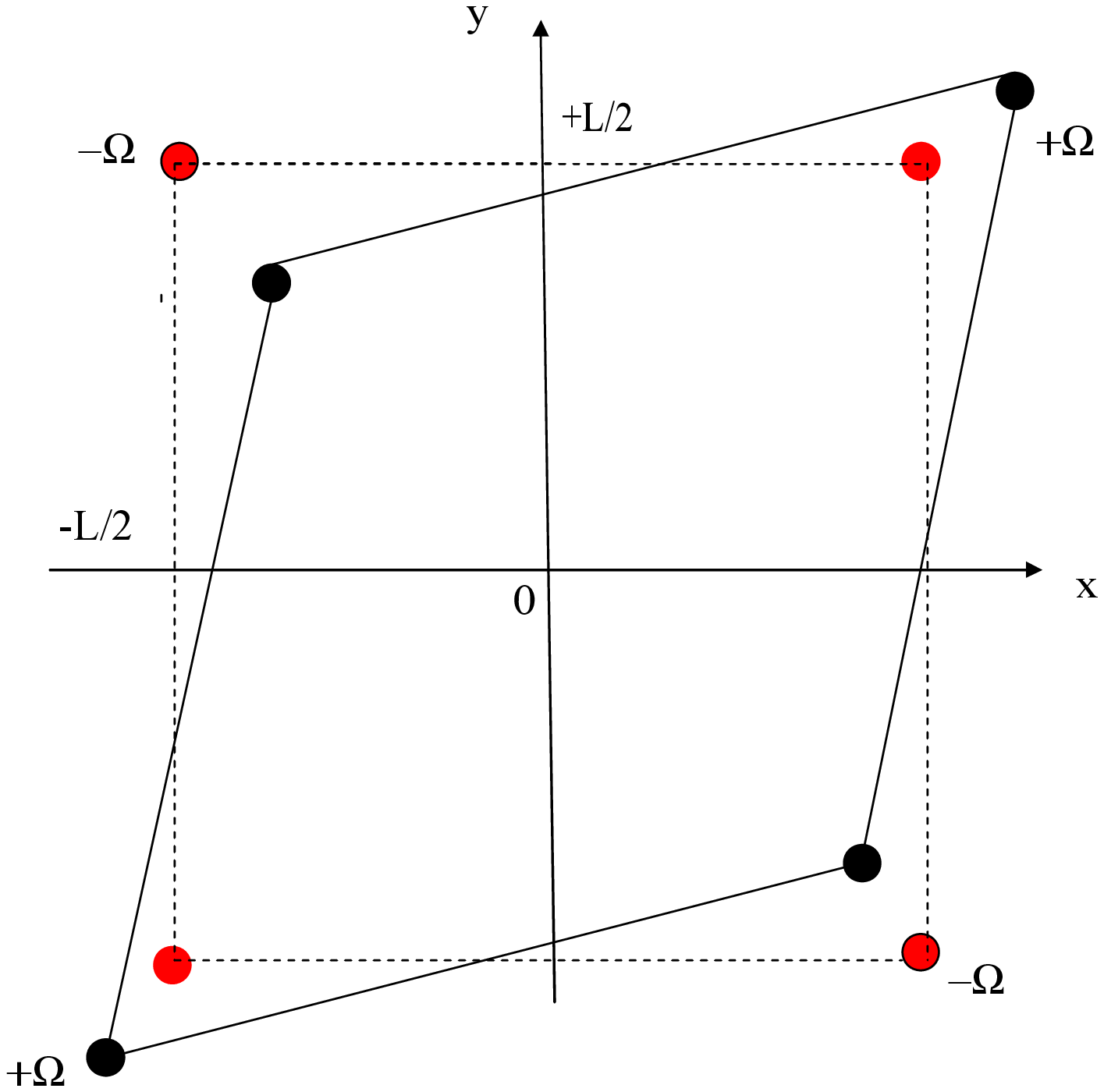}
\end{center}
\caption{Flat SW defect in graphene lattice (left) and its schematic illustration (right) as 5-7-5-7 wedge disclination quadrupole.}
\end{figure}
Actually, this is nothing else than a 5-7-5-7 disclination  quadrupole or, more precisely, the wedge disclination quadrupole (WDQ). The disclination lines with Frank vectors ${\bf \Omega}=\pm\Omega {\bf e_{z}}$ are oriented along the $z$-axis with coordinates $(\pm L/2+\delta, L/2\pm\delta)$ in the positive $xy$ half plane and $(\mp L/2-\delta, -L/2\mp\delta)$ in the negative one. Generally, we consider a rhombus which is  transformed into the square for $\delta=0$. The axes of the rotation of each disclination are not shifted relative to the disclination lines. Otherwise, extra dislocations would be present.

With this understanding, one can describe the SW-phonon scattering in terms of the deformation potential theory. This approach was successful in the description of the phonon scattering by disclination dipoles in dielectric materials~\cite{kras1, kras2} and, more recently, by grain boundaries in graphene~\cite{koles}.
Namely, a phonon mean free path due to WDQ strain fields can be calculated  within the deformation potential approach supposing the deformations are dilatations~\cite{zim}. In this case, an effective perturbation energy  reads
 \begin{equation}
U({\bf r})=\hbar \omega _{\lambda}(q)\gamma_{\lambda }TrE_{ij}(\bf r),
\end{equation}
where $\hbar \omega _{\lambda}(q)$ is the phonon energy with wavevector ${\bf q}$, $TrE_{ij}({\bf r})$ is the trace of the strain tensor caused by the static WDQ,  $\gamma _{\lambda }$ is the Gr\"uneisen constant for a given phonon branch $\lambda $. At chosen in Fig. 1 geometry, the strain matrix for WDQ is known (see, e.g.,~\cite{wit}) and $U({\bf r})$ takes the following form:
\begin{equation}
U({\bf r})=\frac{1}{2}A\biggl(\ln\frac{(x-L_{1}/2)^2+(y-L_{1}/2)^2}{(x+L_{2}/2)^2+(y-L_{2}/2)^2}+\ln\frac{(x+L_{1}/2)^2+(y+L_{1}/2)^2}{(x-L_{2}/2 )^2+(y+L_{2}/2)^2}\biggr),
\end{equation}
where $A=\hbar \omega_{\lambda }(q)\nu \gamma_{\lambda} (1-2\sigma )/(1-\sigma )$, $\nu =\Omega /2\pi $, $\sigma $ is the Poisson constant, $L_1=L+2\delta, L_2=L-2\delta$.
The phonon mean free path is given by
\begin{eqnarray}
l^{-1}_{sw}=n_{sw}\int_{0}^{2\pi}(1-\cos\theta )R(\theta )d\theta
\end{eqnarray}
with $R(\theta )$ being the effective differential scattering radius, $\theta $ the scattering angle, and $n_{sw}$ the areal density of WDQ. Within the Born approximation $R(\theta )$ is written as
\begin{eqnarray}
R(\theta )=\frac{qS^{2}}{2\pi \hbar ^{2}v_{\lambda }^{2}}\overline{|\langle {\bf q}|U({\bf r})|{\bf q^{'}}\rangle|^{2}},
\end{eqnarray}
where $S$ is a projected area, $v_{\lambda }$ is the sound velocity, and the bar denotes an averaging procedure over $\alpha$ which defines an angle between the scattering vector $\bf q - \bf q^{'}$ and the x-axis.
Finally, we obtain the following exact expression:
\begin{eqnarray}
l^{-1}_{sw,\lambda}=4q\nu ^2n_{sw}B^2\Biggl(2\tilde L^2\biggl(J_{0}^{2}(q\tilde L)+J_{1}^{2}(q\tilde L)-
J_{0}(q\tilde L)J_{1}(q\tilde L)/q\tilde L\biggr)- \\ \nonumber
\sum_{n=1}^2 L_n^2\biggl(J_{0}^{2}(\sqrt{2}qL_n)+J_{1}^{2}(\sqrt{2}qL_n)-
J_{0}(\sqrt{2}qL_n)J_{1}(\sqrt{2}qL_n)/\sqrt{2}qL_n\biggr)\Biggr),
\end{eqnarray}
where $J_{i}(z)$ is the Bessel function of the $i$-th kind, $B=\gamma_{\lambda} (1-2\sigma )/(1-\sigma )$, $n_{sw}$ is the areal density of WDQ, $\tilde L=\sqrt{L^2+4\delta^2}$. In graphene lattice with flat SW defect the parameter $\delta$ is small, so that almost quadratic WDQ is realized (see Fig. 1). For this reason, we restrict our further consideration to the case $\delta=0$.

The total phonon mean free path including the most important scatterers reads
\begin{equation}
1/l_{tot,\lambda }=1/l_{0 }+1/l_{sw,\lambda }+1/l_{pd,\lambda }+1/l_{ph-ph,\lambda },
\end{equation}
where $l_{0}$, $l_{pd ,\lambda }$ and $l_{ph-ph,\lambda }$ come from the phonon-rough boundary, phonon-point defect (PD) and phonon-phonon scattering, respectively.
Explicitly,
\begin{equation}
l_{0}^{-1}=\frac{1}{d}\frac{1-p}{1+p},
\end{equation}
where $d$ is the graphene layer size and $p$ is the  specularity parameter that can be chosen to be momentum-independent~\cite{nika1}.
The phonon-PD mean free path takes the form
\begin{equation}
l_{pd,\lambda }^{-1}=\frac{S_{0}\Gamma}{4}\frac{q}{v_{\lambda }^{2}(\omega )}\omega ^{2}_{\lambda}(q),
\end{equation}
where $S_{0}$ is the cross-section area per one atom of the lattice and $\Gamma $ is the measure of the scattering strength. Like in~\cite{aks} we ignore the momentum-conserving three-phonon processes (normal N processes) and restrict our consideration to the resistive umklapp phonon-phonon scattering. In this case,
the mean free path can be written as~\cite{slack,morelli}
\begin{equation}
l_{U,\lambda }^{-1}=B_U\omega ^{2}_{\lambda}(q)(T/\Theta _{\lambda })\exp(-\Theta _{\lambda }/bT),
\end{equation}
with $B_U$ and $b$ being two adjustable constants. Typically, $b\sim 3$ and
$B_U\simeq \hbar \gamma _{\lambda }^2/(\overline {M}\bar v_{\lambda }^3)$,   where
 $\overline {M}$ is the average atomic mass, $\Theta _{\lambda }$ is the Debye temperature, and $\overline {v_{\lambda }}$ is the average sound velocity for the branch $\lambda $. For graphene, this empirical expression was used in~\cite{aks} with $b=3$. Notice that we were managed to fit the results of~\cite{aks} for $\Gamma=10^{-4}$ by using a different value of $b$ ($b=4.5$) and markedly (four times) increasing $B_U$. Alternatively, one could decrease the Debye temperatures. As was discussed in~\cite{morelli}, choosing too high a value for the Debye temperature requires a larger values of $b$ or the coefficient $B_U$. We have used the fitted in such a way parameters in all our calculations.

In order to distinguish between the SW and vacancy phonon scattering we show the corresponding mean free paths in Fig. 2.
\begin{figure} [tbh]
\begin{center}
\includegraphics [width=10.5 cm]{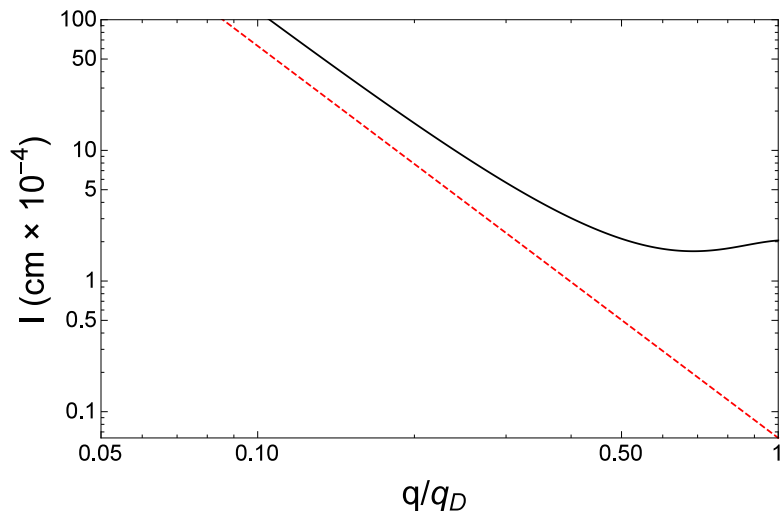}
\end{center}
\caption{The mean free paths of longitudinal phonons $l_{sw,L}$ (solid line) and $l_{pd,L}$ (dashed line) as a function of $q/q_{D}$ with $q_D$ being the Debye wavevector. The parameter set used is: $\gamma _{LA}=2.0$, $\Gamma =0.001$, $L=2.5$\AA, $n_{sw}=2.0\times 10^{12}$ cm$^{-2}$, $\nu =0.16$, $\sigma =0.165$.}
\end{figure}
As is seen, at low wavevectors the behavior is quite similar. With increasing $q$ the situation changes drastically:
$l_{sw}$ is plateaued after some oscillations starting from $q\sim 0.7q_D$. This behavior resembles that found for biaxial wedge disclination dipole~\cite{kras1,kras2}. The extreme points are caused by the Bessel functions where two characteristic lengths exist: the side and the diagonal of the square. $l_{pd}$ shows a steady $q^{-3}$ behavior for all wave vectors.

The total thermal conductivity includes all possible scatterers and dominant phonon branches. We take into account the most important acoustic phonon branches: transverse (TA), longitudinal (LA), and  out-of-plane (ZA). Explicitly, one gets (see, e.g.,~\cite{nika})
$$
\kappa (T) = \frac{1}{4\pi k_{B}T^2h_{eff}}
$$
\begin{equation}
\sum_{\lambda }
\int_0^{q_{max}}(\hbar \omega _{\lambda }(q))^{2}v_{\lambda }(q)l_{tot,\lambda }(q,T)e^{\hbar \omega_{\lambda }(q)/(k_{B}T)}(e^{\hbar \omega_{\lambda }(q)/(k_{B}T)}-1)^{-2}qdq,
\end{equation}
where summation is performed over phonon polarization branches with the dispersion relations $\omega _{\lambda }(q)=qv_{\lambda }$  for $\lambda=LA,TA $, and $\omega _{\lambda }(q)=q^2/2m$ for $\lambda=$ ZA ($m$ is an effective parameter), $k_{B}$ is the Boltzmann
constant, $l_{tot,\lambda }(q,T)$ is the phonon mean free path given by Eq.(6), $h_{eff}$ is the effective graphene layer thickness.

Fig. 3 shows the calculated thermal conductivity in a 5 $\mu$m wide ribbon with vacancies of different concentrations.
\begin{figure} [tbh]
\begin{center}
\includegraphics [width=10.5 cm]{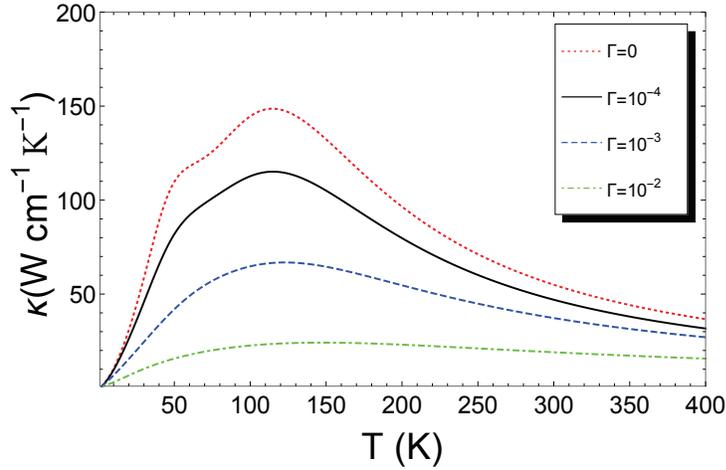}
\end{center}
\caption{Thermal conductivity versus temperature in a 5 $\mu$m wide ribbon with vacancies of different concentrations and $n_{sw}=0$. Used model parameters are:  $\gamma _{LA}=2.0$, $\gamma _{TA}=0.66$,  $\gamma _{ZA}=-1.5$,
$h_{eff}=0.335$ nm, $p=0.7$, $v_{LA}=21.3\times 10^{5}$ cm/s,  $v_{TA}=13.6\times 10^{5}$ cm/s,
$d=5\times 10^{-4}$ cm, $m=100$ s/cm$^{2}$.
 }
\end{figure}
 We use the parameter set typical for graphene and the values of adjustable constants of phonon-phonon umklapp scattering are taken from a fit to the results of~\cite{aks} at $\Gamma=10^{-4}$ ($n_{pd}\sim 2\times 10^{11}$ cm$^{-2}$) which, in turn, are in good agreement with experimental data~\cite{balandin,ghosh}.  As is seen, at low vacancy concentrations the umklapp scattering dominates at room temperatures which agrees with conclusions in~\cite{aks}. At higher concentrations the contribution from vacancy scattering markedly increases, however, the umklapp processes are still of importance. This qualitatively agrees with the results of MD simulations and an analysis made in~\cite{nika}.
Notice  that for pristine GNRs the calculated curve has a false bump in the temperature range of 50-90 K. This follows from the fact that the phenomenological expression (9) for umklapp scattering is not universal and gives good fits in the restricted temperature range. For example, for CdTe the correct region was found to be $0.05\Theta<T<2\Theta$~\cite{slack}.

Evidently, with a proper definition of $\Gamma$, the results shown in Fig. 3 are also valid for isotope scattering  (cf. Ref.~\cite{aks}). A more accurate study of this problem has been recently presented in~\cite{alofi} within the Callaway's theory in its full form (normal processes are included). The authors~\cite{alofi} found a qualitatively similar behavior of the thermal conductivity and made a conclusion that the isotopic effect
on the conductivity is significant in the low-temperature range 50-300 K.

The calculated thermal conductivity in the presence of SW defects is shown in Fig. 4.
\begin{figure} [tbh]
\begin{center}
\includegraphics [width=10.5 cm]{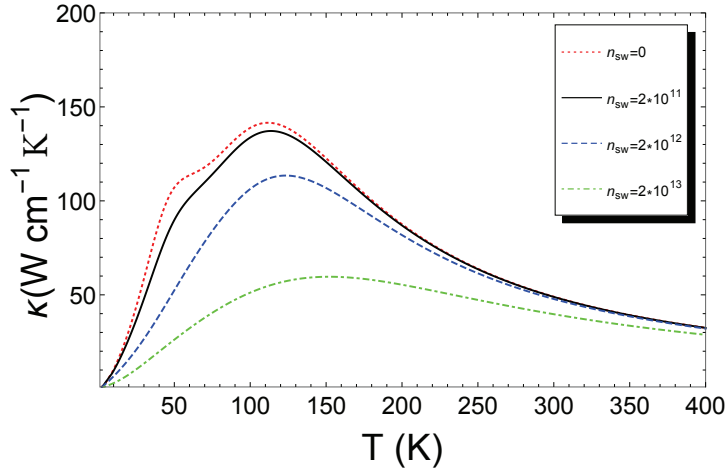}
\end{center}
\caption{Thermal conductivity versus temperature in a 5 $\mu$m wide ribbon with SW defects of different concentrations and $\Gamma=0$.
Used model parameters are the same as in Figs. 2 and 3.}
\end{figure}
As is seen, there is a similar decrease within a wide temperature range though SW defects have less impact on the thermal conductivity in comparison to vacancies with the same concentrations. Qualitatively, this conclusion agrees with the results of MD calculations. However, at room temperature the reduction in the thermal conductivity due to SW defects  is found to be markedly smaller because of the dominant influence of phonon-phonon umklapp scattering. A possible healing of the graphene monovacancies~\cite{rob,zan} and the SW defects~\cite{gur} can additionally reduce the corresponding contributions to the thermal conductivity via decreasing the defect densities, especially at high temperatures.  Notice that a crucial parameter for phonon-SW scattering is a size of the quadrupole: the more size the more reduction in the thermal conductivity takes place.

We have compared the role of SW and vacancy defects in Fig. 5.
\begin{figure} [tbh]
\begin{center}
\includegraphics [width=10.5 cm]{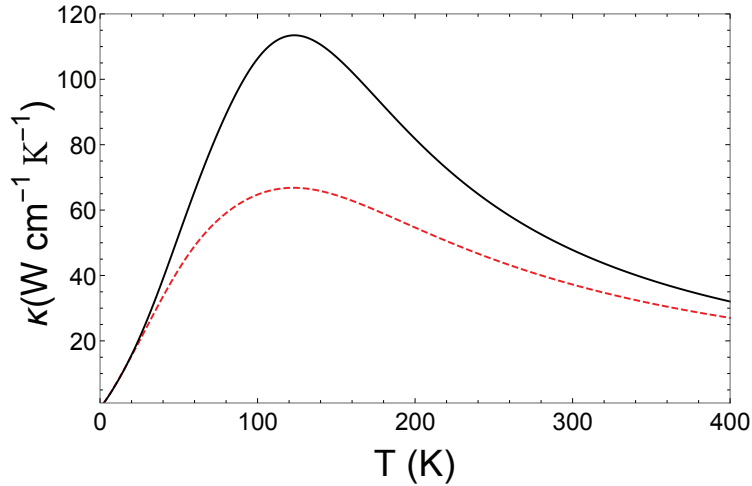}
\end{center}
\caption{Thermal conductivity versus temperature in a 5 $\mu$m wide ribbon with SW defects (solid line) and vacancies (dashed line) at
$n_{sw}=2\times 10^{12}$ cm$^{-2}$,  $\Gamma=10^{-3}$, and the parameter set is the same as in Figs. 2 and 3.}
\end{figure}
As is seen, the SW-induced contribution is less pronounced in comparison to monatomic vacancies of the same concentration. The maximum difference occurs in the region near the thermal conductivity peak ($\sim 120$K). At low temperatures the phonon-rough boundary scattering prevails while at high temperatures the umklapp three-phonon processes are essential. Notice that our analysis confirms the conclusions in~\cite{aks} concerning the role of different phonon branches. Like in~\cite{aks} for both SW and vacancy defects the TA mode has the largest contribution to the thermal conductivity at room temperature whereas the ZA mode is stronger at low temperatures (below the peak).

In conclusion, within the Born approximation we have obtained the exact analytical result for the mean free path due to phonon-SW scattering. This allows us to calculate the corresponding contribution to the thermal conductivity in a wide temperature range and compare it to other scattering mechanisms.
The results demonstrate that SW defects markedly decrease the thermal conductivity below $250$ K. At higher temperatures, the role of three-phonon umklapp scattering becomes dominant. The comparison with vacancy defects shows that the influence of point impurities being qualitatively similar is more pronounced in the same temperature range at equal concentrations. The reason is clearly seen in Fig. 2: the mean three path $l_{sw}$ resembles $l_{pd}$ at low wavevectors whereas at high $q$ the role of phonon-SW scattering diminishes.

Our consideration gives a possibility to analyse heat conduction in graphene nanoribbons with various widths, edge roughness and SW defect concentrations.
In this paper, we have restricted our analysis to the case when Young's modulus does not depend on the concentration of defects. Notice, however, that a recent study in~\cite{nature} shows that the in-plane Young's modulus increases with increasing defect density up to almost twice the initial value for vacancy content of ~0.2$\%$. It is not clear whether something similar can be observed in a graphene lattice with SW defects, which also can be introduced by ion or electron beam irradiation~\cite{kotakovski}. In that case, the mean free paths due to both the SW and three-phonon umklapp scattering would be markedly influenced.

We have used a rather simple phenomenological single-mode  relaxation  time  approach with the smallest possible set of parameters. The more carefull analysis requires inclusion of the three-phonon normal processes, for example, as it has been done in ~\cite{alofi} within the Callaway's theory. Another interesting problem concerns the role of the ripples in graphene. We have considered the flat graphene flake. A wavy shape will induce a  local symmetry breaking of graphene lattice. As a result, the phonon symmetry selection rule will be broken~\cite{nika2}. This, in turn,  leads to the interaction between different phonon branches and, in particular, to the damping of the ZA mode. Thus, the waving should lead to a reduction in the thermal conductivity. This effect is likely to be not significant for small flakes, but should be noticeable when the flake size increases.  The detailed study of this problem still remains open. In our case, the explicit form of a wavy graphene shape should be described in the presence of SW defects, which even complicates the analysis.

\end{document}